\documentclass[%
aip,jcp,amsmath,amssymb, reprint, twocolumn]{revtex4-2}
\usepackage{amsmath}
\usepackage{mathtools}
\usepackage{float}
\usepackage{amsfonts}
\usepackage{amssymb}
\usepackage{dcolumn} 
\usepackage{array}
\newcolumntype{P}[1]{>{\centering\arraybackslash}p{#1}}
\newcolumntype{M}[1]{>{\centering\arraybackslash}m{#1}}
\newcolumntype{C}[1]{>{\centering\arraybackslwash}p{#1}}
\usepackage{float}
\usepackage{psfrag}
\usepackage{tabularx}
\usepackage{stackengine}
\usepackage{amssymb}
\usepackage{mathtools}  
\usepackage{xfrac} 
\usepackage[T1]{fontenc}
\usepackage{graphicx}
\usepackage{diagbox}
\usepackage[caption=false]{subfig}
\usepackage{tikz}
\usetikzlibrary{positioning}
\usetikzlibrary{arrows}
\usetikzlibrary{trees}
\usepackage{amssymb}
\usetikzlibrary{decorations.pathmorphing}
\usetikzlibrary{decorations.markings}
\usetikzlibrary{automata,positioning}
\usepackage{braket}
\usepackage{hyperref}
\usepackage{rotating}
\usepackage{adjustbox}
\usepackage{ragged2e}
\usepackage{simplewick}
\usepackage{simpler-wick}
\usepackage{csquotes}
\usepackage{physics,amsmath}
\usepackage{xcolor}
\usepackage{fancyhdr}
\UseRawInputEncoding
\usepackage[normalem]{ulem}

\usepackage{scalerel}

\begin{document}

\author{Sonaldeep Halder}
\affiliation{ Department of Chemistry,  \\ Indian Institute of Technology Bombay, \\ Powai, Mumbai 400076, India}

\author{Kartikey Anand}
\affiliation{ Department of Chemistry,  \\ Indian Institute of Technology Bombay, \\ Powai, Mumbai 400076, India}

\author{Rahul Maitra}
\email{rmaitra@chem.iitb.ac.in}
\affiliation{ Department of Chemistry,  \\ Indian Institute of Technology Bombay, \\ Powai, Mumbai 400076, India}
\affiliation{Centre of Excellence in Quantum Information, Computing, Science \& Technology, \\ Indian Institute of Technology Bombay, \\ Powai, Mumbai 400076, India}

\title{Construction of Chemistry Inspired Dynamic Ansatz Utilizing Generative Machine Learning}


\begin{abstract}
    Generative machine learning models like the Restricted Boltzmann Machine (RBM) provide a practical approach for ansatz construction within the quantum computing framework. This work introduces a method that efficiently leverages RBM and many-body perturbative measures to build a compact chemistry-inspired ansatz for determining accurate molecular energetics. By training on low-rank determinants derived from an approximate wavefunction, RBM predicts the key high-rank determinants that dominate the ground-state wavefunction. A shallow depth ansatz is constructed to explicitly incorporate these dominant determinants after dynamically decomposing them into low-rank components and applying many-body perturbative measures for further screening. The method requires no additional measurements beyond the initial training phase. Moreover, it incorporates Bayesian hyperparameter optimization for the RBM, ensuring efficient performance with minimal training data during its limited usage. This approach facilitates the efficient computation of molecular properties, paving the way for exploring new chemical phenomena with near-term quantum computers.

\end{abstract}

\maketitle

\section{Introduction}

The utilization of quantum computing
platforms for the evaluation of molecular
energetics and associated properties have
opened up new avenues in the domain of 
electronic structure theory. In this regard, variational
algorithms \cite{peruzzo2014variational,
delgado2021variational, 
grimsley2019adaptive,halder2022dual,
halder2023corrections, 
mondal2023development,feniou2023overlap,zhao2023orbital, tang2021qubit, yordanov2021qubit, ostaszewski2021structure, tkachenko2021correlation, zhang2021adaptive, sim2021adaptive}, which
involve the dynamic construction and
deployment of shallow depth parameterized
ansatzes, have gained widespread
popularity. They are suitable for Noisy
Intermediate-Scale Quantum (NISQ)
\cite{preskill2018quantum} devices that
suffer from limited coherence times, state
preparation and measurement (SPAM) errors,
and poor gate fidelity. However, many of these methods require a substantial number of measurements during the ansatz construction, and the inherent noise in the hardware can substantially affect the final ansatz.
Some methods employ alternate strategies that bypass such measurements, either completely or
partially\cite{halder2024noise, halder2024machine}. A promising method for this
category (developed by some present
authors) combines generative machine
learning (ML) and many-body perturbative
measures to obtain a compact ansatz\cite{halder2024machine}.
However, it involves repeated training and
prediction cycles, increasing the overall 
cost. Additionally, the ML model uses suboptimal hyper-parameters,
leading to a larger requirement for
training data. These limitations underscore the need for more efficient strategies in utilizing generative ML models for ansatz construction.

In this work, we develop a methodology that efficiently utilizes Restricted Boltzmann Machines (RBM)\cite{montufar2018restricted, fischer2012introduction, freund1991unsupervised, tieleman2008training, hinton2006fast, carreira2005contrastive} to generate a shallow depth ansatz. The model is trained using an approximate wavefunction that explicitly involves only low-rank determinants, such as singles and doubles, which span the \textit{primary excitation subspace}\cite{halder2024machine}. This is followed by generating high-rank many-body determinants (such as triples, quadruples, etc., which span the \textit{secondary excitation subspace}) that may have dominant contributions to the ground state molecular wavefunction. These determinants are decomposed on the fly into low-rank operators, followed by a secondary
pruning mechanism based on many-body perturbative measures. It leads to a shallow depth ansatz -- which we call RBM1s-dUCC -- that can be utilized on NISQ platforms to obtain accurate ground state energies. The developed approach limits the need for measurements to the training data acquisition phase, significantly reducing overall measurement costs and minimizing the impact of noise on the entire ansatz construction pipeline\cite{halder2024machine}. It only employs one-step utilization of RBM while incorporating determinants of a particular higher rank. The method also incorporates a Bayesian\cite{snoek2012practical,bergstra2012random, bergstra2013making} based hyperparameter optimization for RBM to ensure it works efficiently during its limited usage. Once RBM1s-dUCC is constructed, it is variationally optimized to obtain ground state energies. The accuracy achieved after this optimization is highly dependent on the starting point\cite{truger2024warm, tilly2022variational, jattana2023improved}. Following a discussion on the development of the method and its associated components, this manuscript also includes a rigorous study regarding the effect of different initial points on the optimization performance. This highlights the presence of local minimum or barren plateaus\cite{larocca2024review, mcclean2018barren, anschuetz2022quantum, qi2023barren, arrasmith2022equivalence, larocca2022diagnosing, cerezo2023does,cybulski2023impact, ragone2023unified, fontana2023adjoint, bremner2009random, gross2009most} in the optimization landscape and helps identify the most effective initial points for the variational optimization of the RBM1s-dUCC ansatz. This method offers an efficient pathway for computing molecular ground state energies on NISQ hardware, particularly for molecular systems where the RBM can effectively learn the correlations from \textit{primary excitation subspace}.

\section{Theory}
In this section, we start by describing briefly the functioning of the Restricted Boltzmann Machine (RBM). It includes key points regarding how RBM can be used to represent many-body wavefunctions and can be extended further to generate an ansatz. Following this, we provide a stepwise method that utilizes RBM and many-body perturbative measures in an efficient way to construct the shallow depth ansatz. In this section, we also discuss the potential limitations of this approach. We then describe the cost associated with our procedure highlighting its efficiency. We end this section with a discussion on hyperparameters that influence the working of RBM and highlight the procedure for obtaining the optimum hyperparameters.

\subsection{Use of Restricted Boltzmann Machine (RBM) to Represent Many-Body Wavefunctions}

Generative neural network models, such as RBM, have been widely used to represent highly entangled quantum states\cite{xia2018quantum,sajjan2022quantum,sajjan2021quantum,sureshbabu2021implementation,sajjan2023imaginary,barrett2022autoregressive, han2019solving, hermann2020deep, pfau2020ab, kessler2021artificial, coe2018machine, hinton2006reducing, melko2019restricted, herzog2023solving}. Beyond tomography\cite{torlai2018neural} and error mitigation\cite{bennewitz2022neural}, these models have been successfully used to construct dynamic ansatzes for molecular systems\cite{halder2024machine}. The key idea behind implementing generative models is that they can effectively encode the diverse contributions of various bases to a given wavefunction. In other words, such models can generate a compact representation of many-body wavefunctions while keeping intact the correlations present within them.

In particular, RBM comprises two layers, a visible and a hidden layer. The visible layer ($v_{i}\in \{0,1\}$) corresponds to observations from the training data. This data consists of various many-body determinants occurring with frequencies dictated by their coefficients in a given wavefunction (which we aim to model). Hence, $v_i$ represents a binary vector representation of a many-body determinant. The method employed to generate such a dataset is described in Section II B. The hidden layer ($h_{i}\in \{0,1\}$) is responsible for capturing the underlying interrelations between different determinants in the given wavefunction. A bias is provided for each visible and hidden unit. A weight matrix is set to establish a connection between the visible and the hidden layers.  RBM aims to find the optimal values of the biases and the weight matrix so that the probability distribution modeled explains the observed data well. RBM constructs a joint probability distribution for the configuration $\{v,h\}$ -

\begin{equation} \label{prob_energy}
    p(v,h)= \frac{1}{z} e^{-E(v,h)}
\end{equation}
where $E(v,h)$ is the Energy function and z is the partition function.
\begin{equation}
     z= \sum_{v,h} e^{-E(v,h)} 
\end{equation}

$E(v,h)$ is parameterized by the biases ($\{b_j\}_j$ and $\{c_i\}_i$) and weights ($W_{ij}$). The parameterized form of $E(v,h)$ is written as -

\begin{equation} \label{score_energy}
    E(v,h) = -\sum_{i=1}^{n}\sum_{j=1}^{m} w_{ij}h_{i}v_{j} -\sum_{i=1}^{n} c_{i}h_{i}
    -\sum_{j=1}^{m} b_{j}v_{j}
\end{equation}
Here, $n$ is the number of hidden nodes and $m$, the number of visible nodes. The RBM is trained by adjusting its weights and biases such that many-body determinants that have higher probabilities in the obtained measurement data are assigned lower energies ($E(v,h)$) and, thus, higher probabilities (Eq. \eqref{prob_energy}. To do that, the log of \textit{likelihood function} ($\mathcal{L}$) is constructed -

\begin{equation}
    \ln\mathcal{L}(\Omega|v) = \ln p(v|{\Omega}) = \ln\frac{1}{z}\sum_{h} e^{-E(v,h)}
\end{equation}
where $\Omega$ are the RBM's parameters. This \textit{likelihood function} is maximized by vanishing its gradient -
\begin{equation}
    \frac{\partial \ln\mathcal{L}(\Omega|v)}{\partial \Omega} = - \sum_{h}p(h|v) \frac{\partial E(v,h)}{\partial \Omega} +  \sum_{v,h}p(v,h) \frac{\partial E(v,h)}{\partial \Omega}
\end{equation}
where $p(h|v)$ is the conditional probability of $h$ given $v$ and is given as -
\begin{equation}
    p(h|v) = \frac{e^{-E(v,h)}}{\sum_{h} e^{-E(v,h)}}
\end{equation}

In practice, direct calculation of the gradient is computationally expensive. In this work, Persistent Contrastive Divergence (PCD)\cite{tieleman2008training} is employed to overcome this. Once the training is complete, the RBM can be used to reconstruct the initial approximate wavefunction in terms of its probability distribution over the many-body basis. In Section IIB, we further explore how the structure of $E(v,h)$ contributes to the generation of new dominant higher-order excitation operators, once the weights ($\{w_{ij}\}_{i,j}$) and biases ($\{c_i\}_i$ and $\{b_j\}_j$) have been optimized through singles and doubles excited determinants.

\subsection{Compact Ansatz Construction through Optimal Utilization of Restricted Boltzmann Machine and Many-Body Perturbative Measures}

The capability of RBM to model a many-body wavefunction can be leveraged to construct a shallow depth ansatz that can provide accurate ground state energies for molecular systems using quantum hardware. RBM can be trained on the measurement data from an approximate wavefunction. Subsequently, this trained model can be used to generate the dominant many-body determinants that ``upgrade" the approximation. An ansatz can then be composed which explicitly produces the dominant determinants suggested by the RBM.
We now provide a stepwise protocol to efficiently generate a compact ansatz using RBM and many-body perturbative measures. Being shallow in depth, this ansatz is highly suitable for NISQ platforms. Additionally, it only requires a single-step utilization of RBM, with measurements performed solely during the generation of training data.

\textbf{Step-1: Approximate wavefunction construction:} A low-level wavefunction is constructed on a quantum device using the shallow disentangled Unitary Coupled Cluster with Singles and Doubles (dUCCSD) ansatz \cite{evangelista2019exact} $\hat{U}_{SD}(\boldsymbol{\theta})$ acting on reference state $\ket{\phi_o}$ (taken to be Hartree Fock (HF) state). The parameters are optimized variationally using the Variational Quantum Eigensolver (VQE)\cite{peruzzo2014variational} framework. A detailed illustration of the quantum circuit used to implement the dUCC ansatz is provided in point S1 of the Supplementary Information. To reduce the ansatz depth, only double excitation operators with associated absolute first order perturbative estimates (M\text{\o}ller-Plesset or MP2 amplitudes) above a threshold (set to be $10^{-05}$ in this work) are considered. No pruning is done for singles excitation operators. The excitations are arranged in descending order based on their absolute MP2 values, with the highest acting first on the reference state.
\begin{equation}
\begin{split}
    \ket{\Psi_{SD}} &= \hat{U}_{SD}(\theta^{opt})\ket{\phi_o}\\
    &=e^{\theta_{S_{n}}^{o}\hat{\kappa}_{S_{n}}}e^{\theta_{S_{n-1}}^{o}\hat{\kappa}_{S_{n-1}}} \dots e^{\theta_{D_2}^{o}\hat{\kappa}_{D_2}}e^{\theta_{D_1}^{o}\hat{\kappa}_{D_1}}\ket{\phi_o}
\end{split}
\end{equation}
$\hat{\kappa}_S$ and $\hat{\kappa}_D$ denote anti-hermitian singles and doubles excitation operators, respectively, and $\theta_S^o$ and $\theta_D^o$ are the optimized parameters. 
The doubles excitation operators (having non-zero MP2 values) act first, followed by 
singles (with zero value at MP2 level) to maintain the descending order.

\textbf{Step-2: Measurements for binary vector data acquisition:} The m-qubit approximate state ($\ket{\Psi_{SD}}$) generated by $\hat{U}_{SD}(\theta^{opt})$ acting on the reference HF state is measured in the computational basis. The measurement data provides the probabilities of various many-body determinants that span the \textit{primary excitation subspace} (that is, the determinants having direct correspondence to the excitations taken in $\hat{U}_{SD}$). A binary-vector dataset is prepared where each vector represents a determinant in the \textit{primary excitation subspace} and occurs with the frequency governed by its associated probability in $\ket{\Psi_{SD}}$
\begin{equation}
\begin{split}
    \ket{\Psi_{SD}}\xrightarrow[\text{Collect Data}]{\text{Measure}} &\{P_{S_{n}},\ket{\phi_{S_{n}}}; P_{S_{n-1}},\ket{\phi_{S_{n-1}}}; \\
    &  \dots P_{D_2},\ket{\phi_{D_2}};
    P_{D_1},\ket{\phi_{D_1}}\}
\end{split}
\end{equation}
\begin{equation}
    \ket{\phi_{\mu}} \in \{0,1\}^{n}
\end{equation}
$P_S$ and $P_D$ represent the probabilities with which the singly excited determinants $\ket{\phi_{\mu = S}}$ and doubly excited determinants $\ket{\phi_{\mu = D}}$ occur after measurement. This dataset is normalized to have the sum of probabilities equal to one. HF state is not involved in training as it has a large probability and may bias the model. Once trained, the RBM can model the probability distribution of various basis functions ($\{\ket{\phi_S}\}, \ket{\phi_D}\}$) in $\ket{\Psi_{SD}}$. Corresponding to each possible combination of visible layer and hidden layer ($v \in \{0,1\}^m$ and $h \in \{0,1\}^n$), a probability $p(v,h)$ is assigned (Eq. \eqref{prob_energy},\eqref{score_energy}). Some of the configurations of the visible layer represent the determinants present in the training dataset. The weights and biases ($\{w_{i,j}\}_{i,j}$, $\{b_{j}\}_j$, $\{c_{i}\}_i$) used to parameterize the probability (Eq. \eqref{score_energy}) are adjusted during training (using log-likelihood function and PCD) such that the probability distribution of different determinants in the training data can be modeled using $p(v,h)$. In essence, this step allows the RBM to capture and learn the underlying patterns within the determinants of the \textit{primary excitation subspace}.

\textbf{Step-3: Wavefunction expansion via RBM generation:} The trained RBM is used to generate a batch of determinants, which contain new higher-rank excited determinants (which form the \textit{secondary excitation subspace}) along with the ones already included in the training. The new ones of immediate higher rank, such as triples ($\ket{\phi_T}$) 
(when the initial ansatz is trained with SD), are filtered out from this batch, given they appear at least more than once. Their corresponding excitation operators, in the exponential form, are added to the operator sequence of $\hat{U}_{SD}$. This gives rise to a new ansatz, say $\hat{U}_{ML}$.
\begin{equation}
\begin{split}
    \{P_{S_{n}},\ket{\phi_{S_{n}}}; \dots
    P_{D_1},\ket{\phi_{D_1}}\} \xrightarrow[\text{Generate}]{\text{Train RBM}} \{&\ket{\phi_{S_{n}}}, \dots \ket{\phi_{D_{1}}}, \\
    &\ket{\phi_{T_1}}, \ket{\phi_{T_2}}, \dots \}
\end{split}
\end{equation}
\begin{equation}
    \hat{U}_{ML} (\boldsymbol{\theta}) = \dots e^{\theta_{T_2}\hat{\kappa}_{T_2}}e^{\theta_{T_1}\hat{\kappa}_{T_1}}e^{\theta_{S_{n}}\hat{\kappa}_{S_{n}}}\dots e^{\theta_{D_1}\hat{\kappa}_{D_1}}
\end{equation}
Here, $\hat{\kappa}_{T_1}, \hat{\kappa}_{T_2}, \dots$ represent anti-Hermitian triples excitation operators corresponding to excited determinants $\ket{\phi_{T_1}}, \ket{\phi_{T_2}}\dots$. A mathematical description of how a higher-order excited determinant can be generated, despite the model being trained only on singles and doubles determinants, is as follows: as outlined in Step 2, the training process adjusts the weights and biases ($\{w_{i,j}\}$, $\{b_{j}\}_j$, $\{c_{i}\}_i$) to maximize the likelihood function (or its logarithm) for visible configurations corresponding to singly and doubly excited determinants in the training data. For a visible configuration representing a triply excited determinant ($v_T$), the log-likelihood is given as:

\begin{equation} \label{loglikelihood-triples}
    \ln\mathcal{L}(\Omega^{opt}|v_T) =  \ln\frac{1}{z}\sum_{h} e^{-E(v_T,h)}
\end{equation}

\begin{equation} \label{energyparam-triples}
E(v_T,h) = -\sum_{i=1}^{n}\sum_{j=1}^{m} w_{ij}^{opt}h_{i}v_{T,j} -\sum_{i=1}^{n} c_{i}^{opt}h_{i}
    -\sum_{j=1}^{m} b_{j}^{opt}v_{T,j}
\end{equation}

In Eqs. \eqref{loglikelihood-triples} and \eqref{energyparam-triples}, superscript $opt$ represents optimized weights and biases obtained through training over singles and doubles determinants. Although the likelihood of $v_T$ was never explicitly maximized while training, it can gain value through the action of hidden units and the optimized weights (see the structure of Eq. \eqref{energyparam-triples}). Thus, the patterns learned from singles and doubles (encoded within optimized weights and biases) can be extrapolated with the help of the hidden layer to obtain higher-order determinants.

\begin{figure*}[!ht]
     \centering
 \includegraphics[width=0.9\textwidth]{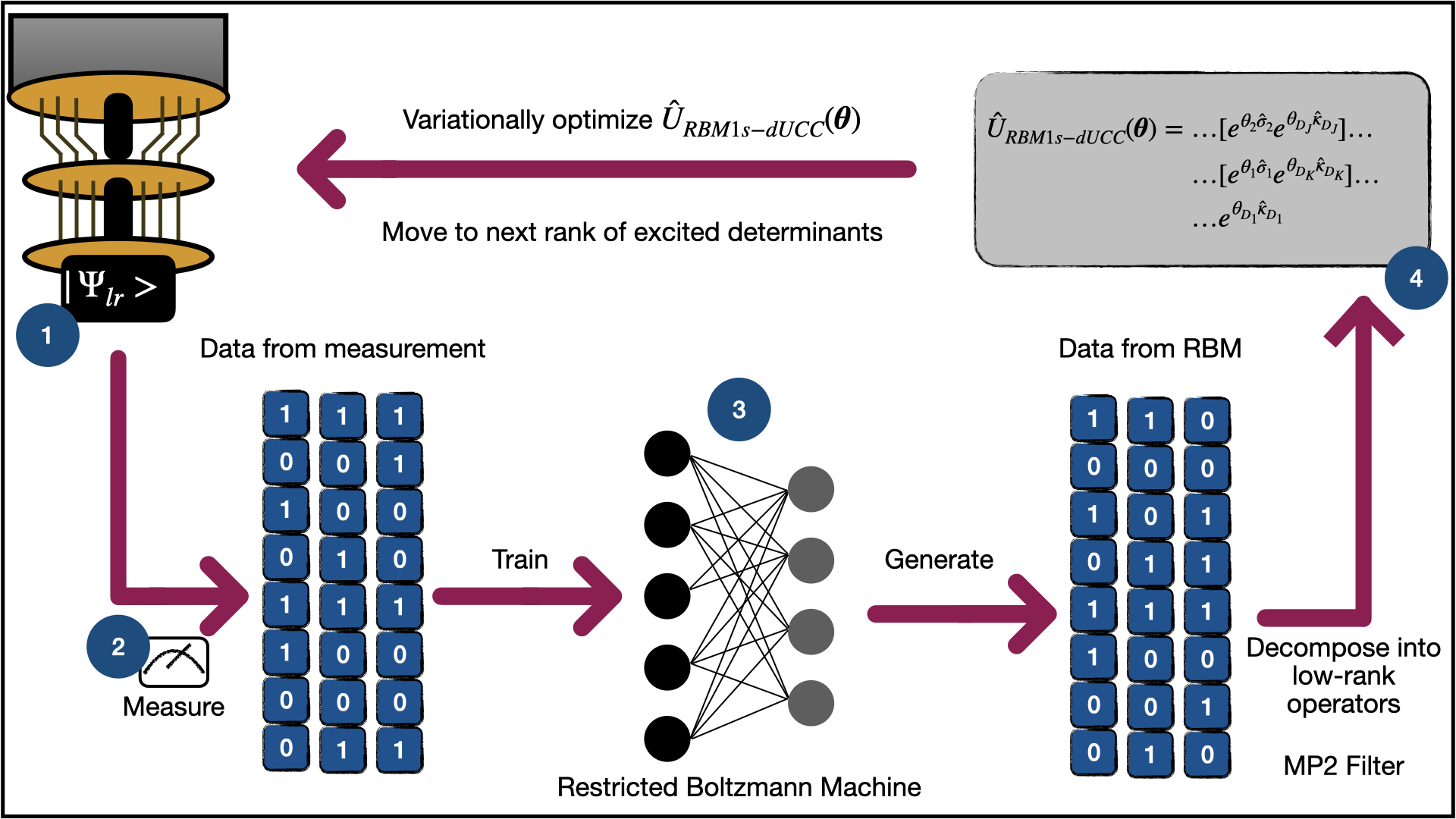}
 \caption{An illustration of Steps 1-4. $\ket{\Psi_{lr}}$ represents an approximate function containing only low-rank determinants explicitly. Initially, this equals $\ket{\Psi_{SD}}$. After completing one cycle, this state now explicitly contains one-rank higher determinants. Continuing the cycle leads to the explicit inclusion of even higher-rank dominant determinants via recursive expansion of the secondary excitation subspace.}
     \label{fig:illustrate_rbm}
 \end{figure*}

\textbf{Step-4: Construction of Ansatz via low-rank decomposition:} Instead of directly adding high-rank excitation operators as for $\hat{U}_{ML}$, they are decomposed 
on-the-fly into \textit{scatterers} $\hat{\sigma}$ \cite{das2005quantum, halder2022dual, halder2023corrections} (which are two-body operators with effective hole-particle excitation rank of one; see S2 of Supporting Information) and a low-rank excitation operator (such as doubles), which is already present in the ansatz. 
\begin{equation} \label{commut_trip}
\begin{split}
    &[\hat{\sigma}_1, \hat{\kappa}_{D_{K}}] \longrightarrow \hat{\kappa}_{T_{1}} \\
    &[\hat{\sigma}_2, \hat{\kappa}_{D_{J}}] \longrightarrow \hat{\kappa_{T_2}}
\end{split}
\end{equation}
More explicitly, each connected triples operator $\kappa_{ijk}^{abc}$ is decomposed as $\hat{\kappa}_{ijk}^{abc} \rightarrow [\hat{\sigma}_{ij}^{al},\hat{\kappa}_{lk}^{bc}]$ or $[\hat{\sigma}_{id}^{ab},\hat{\kappa}_{jk}^{dc}]$ ($\{i,j,k,l\}$ are occupied orbital labels and $\{a,b,c,d\}$ are unoccupied orbital labels), where $\hat{\kappa}_{lk}^{bc}$ or $\hat{\kappa}_{jk}^{dc}$ has already been included in the \textit{primary excitation subspace}. The previous works by Maitra et. al. \cite{maitra2017coupled, tribedi2020formulation, halder2022dual, mondal2023development, halder2024noise} provide a rigorous exploration of such decomposition both in classical and quantum computing. This decomposition ensures the ansatz depth remains low as compared to the direct incorporation of high-rank excitation operators \cite{halder2022dual, halder2023corrections}. The number of Pauli terms obtained after the Jordan-Wigner (JW) transformation of a K-body excitation operator scales as $ \mathcal{O}(2^{2K-1})$. If we decompose such a K-body operator into a commutator of several two-body operators, the scaling goes a $\mathcal{O}(P_F \times 2^3)$, $P_F$ being a prefactor dependent on the number of two-body operators present in the decomposition. For example, if we directly implement a triples excitation operator, it would require $32$ Pauli strings, whereas its decomposition into two two-body operators would require $2\times8=16$ Pauli strings. This reduction directly corresponds to the lowering of gate depth. Moreover, only those high-rank excitation operators are included whose embedded \textit{scatterers} (Eq. \eqref{commut_trip}) have absolute MP2 values above a threshold (chosen as $10^{-5}$ here).  This additional filtration layer ensures only the most dominant excitation operators are added. Instances may arise where distinct \textit{scatterers}, with MP2 values above this threshold, may combine with different low-rank operators already present in the ansatz to produce the same high-rank excitation operator. When this occurs, the \textit{scatterer} with the largest absolute MP2 value is selected, ensuring a unique pathway to achieve the desired high-rank operator\cite{halder2024noise}. The resulting ansatz, which we call RBM1s-dUCC and denoted as $\hat{U}_{RBM1s-dUCC}$, is optimized using VQE. See point S1 of Supplementary Information for the circuit implementation of $\hat{U}_{RBM1s-dUCC}$.
\begin{equation}\label{final_ansatz}
\begin{split}
    \hat{U}_{RBM1s-dUCC} (\boldsymbol{\theta}) = & e^{\theta_{S_{n}}\hat{\kappa}_{S_{n}}} \dots
    [e^{\theta_{2}\hat{\sigma}_{2}}
    e^{\theta_{D_J}\hat{\kappa}_{D_J}}] \dots \\
    &[e^{\theta_{1}\hat{\sigma}_{1}}
    e^{\theta_{D_K}\hat{\kappa}_{D_K}}]
    \dots e^{\theta_{D_1}\hat{\kappa}_{D_1}}
\end{split}
\end{equation}
\begin{equation} \label{expect_min}
    E^{'}= \displaystyle{\min_{\boldsymbol{\theta}} \bra{\phi_0}\hat{U}^{\dag}_{RBM1s-dUCC}(\boldsymbol{\theta})\hat{H}\hat{U}_{RBM1s-dUCC} (\boldsymbol{\theta})\ket{\phi_o}}
\end{equation}
It must be noted in Eq. \eqref{final_ansatz} that not all $\kappa_{D}$ necessarily participate in generating dominant high-rank operators. Moreover, the single excitation operators are at the end according to the ordering set during Step 1. This marks the end of the method, and the resulting minimum value of the expectation term in Eq.(\ref{expect_min}) is the approximate target ground state energy ($E^{'}$). The procedure described in steps 1-4 is illustrated in Fig. \ref{fig:illustrate_rbm}

By completing steps 1-4, the rank of the excited determinants that the initial ansatz can explicitly generate increases by one. Starting with an ansatz that explicitly generates singly and doubly excited determinants, these steps lead to a new ansatz capable of explicitly generating dominant triply excited determinants. We call this new ansatz RBM1s-dUCCSD$T_{S}$. RBM1s signifies that the model is used only for a single step. dUCCSD$T_{S}$ denotes that this ansatz generates singly, doubly, and triply excited determinants explicitly. The subscript $S$ in $T$ highlights that the triples operators are derived through their low-rank decomposition using \textit{scatterers}. An alternative perspective on this procedure is considering the singles and doubles excitation operators as defining the \textit{ Primary Excitation Subspace}. After training on determinants from this subspace, the RBM generates higher-rank determinants that form the \textit{Secondary Excitation Subspace}. This procedure is recursive, with the expansion into the \textit{Secondary Excitation Subspace} occurring incrementally, one excitation rank at a time. Incorporating the higher rank determinants explicitly into the ansatz results in an \enquote{expanded} wavefunction, which captures more correlation. In approaches like Neural Quantum State Tomography (NQST) \cite{torlai2018neural} and quantum machine learning techniques employing Restricted Boltzmann Machines (RBMs) \cite{xia2018quantum, sajjan2022quantum, sajjan2021quantum, sureshbabu2021implementation, sajjan2023imaginary}, the machine learning model must fully encode the wavefunction, including its phase. NQST accomplishes this using two separate RBMs, while the latter employs a three-layer RBM architecture. In contrast, this work utilizes a traditional RBM (described in Section IIA) to capture the dominant high-rank excitations. The complete wavefunction, along with its phase, is then constructed on quantum hardware using the final ansatz ($\hat{U}_{RBM1s-dUCC}$).

The structure RBM1s-dUCCSD$T_{S}$ that includes high-rank correlation
through its decomposition into lower rank operators is reminiscent of double-exponential 
(factorized) coupled cluster ansatz \cite{maitra2017coupled} and its 
unitarized variant \cite{halder2022dual}. Such structure of the  waveoperator in terms of two-body operators is essential towards 
the exactness of the wavefunction for which the variational minimum analytically 
satisfies the contracted Schrodinger equation (CSE)\cite{mazziotti2004exactness} -- a necessary and 
sufficient condition for the wavefunction to satisfy the Schrodinger equation. 
Contrarily, the wavefunctions generated from a single exponential with 
generalized two-body operators\cite{nooijen2000can, nakatsuji1976equation} does not 
satisfy CSE\cite{mazziotti2004exactness}. The CSE and its Anti-Hermitian variant (ACSE) lead to the 
direct determination of energy and two-electron reduced density matrices of 
many-electron molecules\cite{mazziotti2006anti, mazziotti2007anti} and is 
shown to be more expressive and efficient than the UCC ansatz. While ACSE bypasses the explicit construction of the N-electron wavefunction, our ansatz generates the many-body wavefunction without explicit referencing to the reduced density matrices. Moreover, the present approach makes use of generative machine learning for ansatz construction.
ACSE has been implemented both on quantum
simulators and quantum devices\cite{smart2021quantum, smart2022many, smart2022resolving, wang2023quantum, benavides2023quantum} at the cost of shallow quantum circuits.

Using variationally optimized RBM1s-dUCCSD$T_{S}$ as the initial state, repeating steps 2-4 will then include quadruple excitation operators $\hat{\kappa}_{Q}$ (RBM1s-dUCCSD$T_SQ_S$). Such an inclusion would require a combination of two \textit{scatterers}.
\begin{equation}
    [\hat{\sigma}_1^{'},[\hat{\sigma}_1, \hat{\kappa}_{D_{K}}]] \longrightarrow \hat{\kappa}_{Q_{1}}
\end{equation}
The inner commutator is responsible for generating a triples excitation operator. Another suitable \textit{scatterer} combines with this to produce a quadruple excitation operator. Starting from $U_{SD}$ and then continuing this process (steps 2-4), it is possible to include excited determinants of any arbitrary rank explicitly. However, it must be done in steps as a high-rank excitation operator is built using a low-rank excitation operator already present in the ansatz. That is, to incorporate quadruples, the triples excitation operators (in their decomposed form) must already be present. In general, only those $k-$body excitations are included from the batch generated using RBM, which structurally subsumes a dominant $(k-1)$-body excitation that is explicitly or 
implicitly parametrized through $U_{ML}$ in its previous step.

Another point to note is that Step-1 and Step-4 uses MP2 values as additional filtering criteria. Cases may arise, such as in highly correlated molecular systems, where these perturbative measures may become unreliable. However, the conservative threshold applied in Steps 1 and 4 ($10^{-5}$) ensures that the filtering remains resilient to fluctuations in MP2 values. The ordering of excitation operators during the construction of $\hat{U}_{SD}$ in Step 1 may change, but it still contains the same set of excitations ultimately used for RBM training. The effect of such reordering can form the subject for future explorations. In the molecular systems we examined, utilizing MP2 measures resulted in accurate evaluation of molecular energetics. The RBM's capacity to generate higher-rank determinants may be limited if the \textit{primary excitation subspace} fails to capture sufficient correlation patterns. However, since double excitations appear at the lowest order in the perturbative series, a \textit{primary subspace} composed of singles and doubles excitations is generally expected to capture correlations with "qualitative" accuracy, even in highly correlated systems. The RBM1s-dUCC is a dynamic ansatz derived from the dUCC framework, inheriting its correlation-capturing capabilities without exceeding them on its own. For a detailed comparison of dUCC-based methods with alternative approaches, such as subspace expansion, quantum imaginary time evolution, and quantum Monte Carlo, readers can refer to \cite{cao2019quantum, bauer2020quantum, motta2022emerging}.

\subsection{Cost Associated with the Construction of RBM1s-dUCC}

The first step of our procedure, delineated in Section IIB, requires the VQE optimization using a dUCCSD ansatz. The gate complexity for this step using JW mapping is upper bounded by $\mathcal{O}(n_{qubits}^5)$, $n_{qubits}$ being the number of qubits\cite{romero2018strategies}. However, the MP2 pruning used on the doubles excitation operator of dUCCSD reduces this gate complexity in practice. Once the dUCCSD ansatz parameters are variationally optimized, the resulting wavefunction is measured to obtain the training data (Step 2 of Section IIB). The sampling space for such a measurement encompasses all possible many-body determinants produced by the action of the dUCCSD ansatz on the Hartree-Fock (HF) reference state. Among these, the dominant determinants -- those with the largest modulus squared coefficients -- are primarily associated with the single and double excitation operators included in the dUCCSD ansatz, whose number scales as  $\mathcal{O}(n_{qubits}^4)$ (under JW mapping where number of qubits equal number of spin orbitals, given no additional qubit reduction techniques are applied). Consequently, as the system size increases, the number of measurements must be adjusted accordingly. Once the training data is obtained, the RBM comes into play. Training a classical RBM using PCD\cite{scikit-learn} requires a time complexity of $\mathcal{O}(d^2)$ where $d$ is approximately equal to the number of visible nodes ($m$) or the number of hidden nodes ($n$), that is, $d \sim m \sim n$. The number of visible nodes corresponds to the number of qubits ($m = n_{qubits}$). Additionally, as numerically observed in section IIIA, the number of optimum hidden nodes comes out as a multiple of the number of visible nodes. Thus, the time complexity for the RBM training can be approximately written as $ \approx \mathcal{O}(m^2) \approx \mathcal{O}(n_{qubits}^2)$.
Generating dominant determinants using RBM is a variant of selected configuration interaction. As elucidated by Herzog et al.\cite{herzog2023solving}, the bottleneck for such generative approaches arises from verifying whether the generated excited determinants are already incorporated \enquote{within} the ansatz at a given step. This scales as O($N_{det}^2$), with $N_{det}$ being the number of determinants already incorporated. For methods\cite{halder2024machine}, which uses $N_{cyc} (> 1)$ number of cycles of training and prediction to move on to the next rank of excited determinants, this scales as $\mathcal{O}(N_{cyc}N_{det}^2)$. The proposed strategy uses RBM only once (step 3) to generate the next rank of excited determinants. Hence, in this case, $N_{cyc}=1$ and the method scales as $\mathcal{O}(N_{det}^2)$. As the system size increases, the complexity does not necessarily approach that of Full Configuration Interaction (FCI), as $N_{det}$ represents the set of \textit{dominant} configurations. In practice, $N_{det}$ grows sub-exponentially with system size, even for highly correlated systems. Apart from the cost associated with the generative procedure of RBM, the sampling cost of the Hamiltonian expectation value also comes into play. Once the RBM1s-dUCC ansatz is constructed, VQE is run to optimize the parameters that require evaluations of Hamiltonian through statistical sampling\cite{romero2018strategies}. Each term of the molecular Hamiltonian, after being converted into a sum of Pauli strings, requires a repeated measurement of $\mathcal{O}(\frac{1}{\epsilon^2})$, $\epsilon$ being the required precision in the expectation value. More efficient methods of VQE\cite{wang2019accelerated} have been developed, and its incorporation within the present approach could form the subject of future research. One downside of the protocol delineated in Steps 1 to 4 of Section IIB is that the RBM may not capture all possible dominant determinants in a single generation step. Thus, using optimum hyperparameters becomes important to extract the maximum possible advantage of RBM.

\subsection{Description of RBM Hyperparameters and its Impact on the Construction of RBM1s-dUCC ansatz}
The architecture of RBM comprises a visible and a hidden layer. Each layer contains a certain number of nodes\cite{scikit-learn}. The visible layer represents observations from the training data derived by measuring an approximate wavefunction. In contrast, the hidden layer captures underlying patterns within this data. The number of nodes in the hidden layer is a critical hyperparameter that must be carefully selected to optimally capture the patterns within the training set. Once the model learns the pattern, it generates new excited determinants (in binary vectors) using Gibbs sampling\cite{gilks1995markov,scikit-learn}. The number of Gibbs sampling steps is another critical hyperparameter. Increasing the number of steps makes the generated data more closely resemble the training data. However, it's essential to select an optimal number of steps to generate new determinants that, while distinct from the training data, still retain the learned patterns.
Additionally, learning rate and batch size are key hyperparameters for RBM. The learning rate dictates the extent of weight adjustments during training; a higher rate speeds up training but may cause instability. The batch size determines the number of training examples used per iteration to update the model's weights.
 Traditional methods like random search and grid search for hyperparameter optimization have limitations\cite{claesen2015hyperparameter, bergstra2011algorithms}. Random search requires numerous evaluations to find optimal configurations, while grid search faces the curse of dimensionality. The Bayesian approach\cite{bergstra2012random, bergstra2013making, snoek2012practical} offers a more efficient alternative for hyperparameter optimization. In point S3 of the Supporting Information, we provide a theoretical description of the Tree-structured Parzen Estimator (TPE) algorithm\cite{bergstra2011algorithms} (a type of Bayesian optimization) that is used in this work.

\section{Results and Discussions}
\subsection{General Considerations for Numerical Demonstrations}
All the subsequent calculations have been performed using an STO-3G basis. The required orbitals and integrals were obtained through PySCF\cite{sun2018pyscf}, which were used for the molecular Hamiltonians and MP2 values. All required quantum computing components for executing Steps 1-4 have been obtained using Qiskit \cite{Qiskit}. JW mapping has been used to convert second quantized operators to qubit operators. The RBM is constructed using the Scikit-learn package\cite{scikit-learn}. For all the molecular systems studied, the core orbitals have been frozen. The hyperparameters have been optimized using the Optuna\cite{optuna_2019} framework for the molecule of $CH_2$ ($r=1.75 \times r_{eq}; \quad r_{eq}=$1.109\AA, $\angle H-C-H=102.400^o$) which is the last molecular geometry studied in Fig. \ref{fig:energy_gates}. The optimum hyperparameters were selected as (the number of hidden nodes, number of Gibbs sampling, learning rate, batch size) = ($23$, $20$, $0.00198459$, $90$). The same hyperparameters have been used across all other molecular systems. One can choose to perform separate hyperparameter optimization for each molecular system; however, as seen in Fig. \ref{fig:energy_gates}, using the same hyperparameters gives accurate results throughout all studied molecular systems. The RBM has been trained using $10000$ determinants. For the generation step, $10000$ determinants were sampled using Gibbs sampling (this is substantially lower than $500000$ determinants used for training and generation in Ref. \cite{halder2024machine}). The classical optimizer Conjugate Gradient (CG) is used to minimize energy expectation values.

\subsection{Impact of Initial Point on the Accuracy}

\begin{figure*}[!ht]
    \centering
\includegraphics[width=0.9\textwidth]{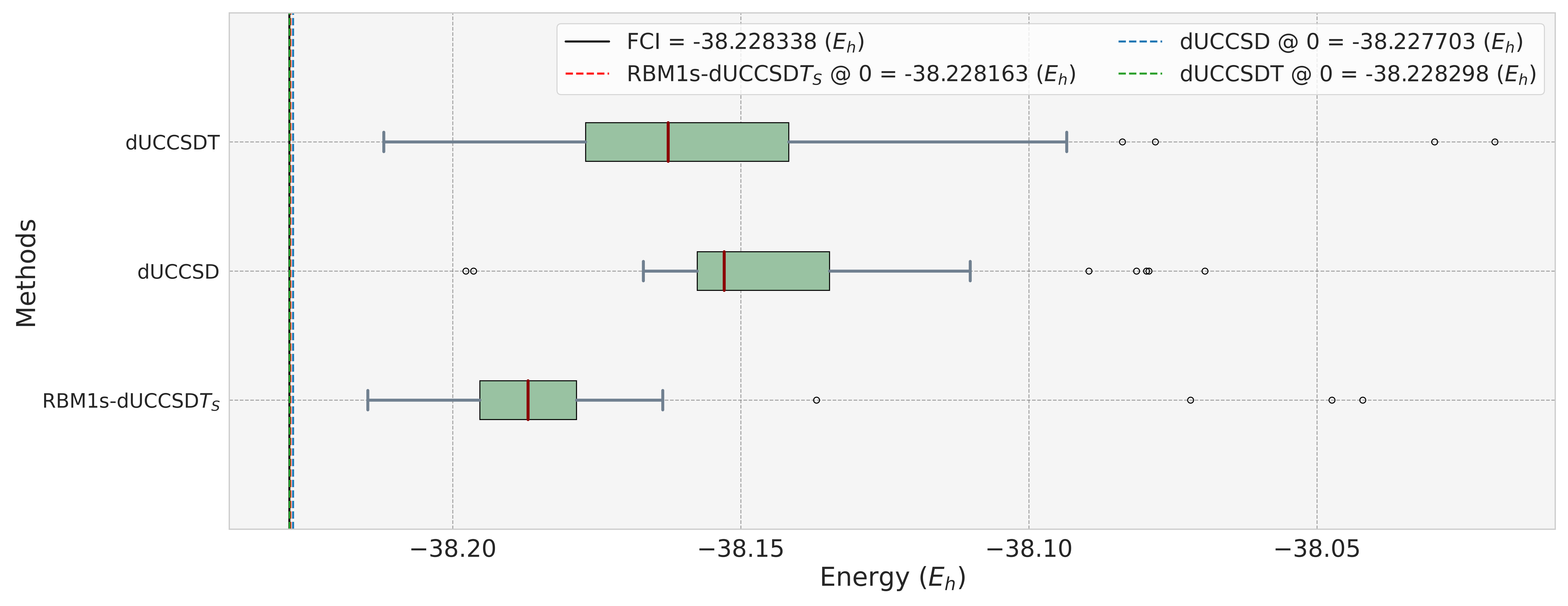}
\caption{Box and whiskers plot representing the distribution of converged energies using dUCCSDT, dUCCSD, and RBM1s-dUCCSD$T_S$ ansatz points for $CH_2$ ($r=1.75 \times r_{eq}; \quad r_{eq}=$1.109\AA, $\angle C-H-C=102.400^o$). The energies obtained by initializing parameters with zero (represented as $@ 0$) and the Full Configuration Interaction (FCI) are also provided.}
    \label{fig:bandw}
\end{figure*}

Before implementing the RBM1s-dUCC ansatz for various molecular systems, it's crucial to establish effective starting points for variational optimization. A key advantage of dUCC-based ansatzes is using MP2 values as starting points. Another common approach is to start optimization from zero. Our studies found that both methods yield nearly identical accuracies (see Supporting Information S4). However, it is also worthwhile to explore how optimization performs when starting with a random set of parameters. This will highlight the presence of local minimum or barren plateaus\cite{larocca2024review, mcclean2018barren, anschuetz2022quantum, qi2023barren, arrasmith2022equivalence, larocca2022diagnosing, cerezo2023does,cybulski2023impact, ragone2023unified, fontana2023adjoint, bremner2009random, gross2009most} in the optimization landscape and helps identify the most effective initial points for the variational optimization. The box and whiskers plot in Fig. \ref{fig:bandw} illustrates the impact of initializing variational optimization with random parameters for the dUCC, dUCCSDT, and RBM1s-dUCCSD$T_S$ ansatzes. For each ansatz, fifty parameter sets were generated, with each parameter in a set drawn from a random uniform distribution between $[−2\pi,2\pi)$, reflecting the periodic nature of the $R_Z (\theta)$ gate in the dUCC ansatz. The resulting energies after optimization using CG (maximum iterations set to $100$) give rise to a distribution. The box and whiskers plot (shown in Fig. \ref{fig:bandw}) is a standardized way of displaying any distribution of data based on a five-point summary: minimum, first quartile (Q1), median (Q2), third quartile (Q3), and maximum. The whiskers at the extreme ends represent the minimum and maximum values of the distribution, the left edge of the box marks Q1, the vertical line inside the box represents Q2, and the end of the box gives Q3. The whiskers, however, do not extend beyond $1.5 \times$ \textit{Inter-Quartile-Range} (represents the distance between Q1 and Q3). The points which lie beyond the maximum extension of the whiskers are represented as circular points and can be treated as outliers. From Fig. \ref{fig:bandw}, it can be observed that the median energies for dUCCSD, dUCCSDT, and RBM1s-dUCCSD$T_S$ lie well above the corresponding energies obtained by initializing parameters with zero. This indicates that the optimization landscapes for the studied ansatzes have a \textit{narrow gorge} around the \enquote{good} minimum. Here, \enquote{good} does not mean a global minimum but where the accuracy is within the desired order. If one starts from parameters initialized at zero or MP2 values, the energy falls within this gorge, which is further iteratively optimized. While the number of iterations in Fig. \ref{fig:bandw} is capped at 100, this is sufficient for the energy to converge to a \enquote{good} value given the initial point lies within the gorge. Starting from random initial parameters, the optimization leads to a poor local minimum or gets trapped in a region of barren plateau. 
Another point to observe is that the median for dUCCSDT has a higher value than RBM1s-dUCCSD$T_S$. However, if we start from initial parameters set to zero, the former gives a lower converged value than the latter (they differ by $\approx10^{-4}$ Hartree). This indicates that although the expressibility\cite{sim2019expressibility} of dUCCSDT and RBM1s-dUCCSD$T_S$ are similar, the larger number of parameters in the case of the former gives it more directions to get trapped when initialized from a random set. In general, from Fig. \ref{fig:bandw}, it can be ascribed that using random initial points during the optimization of RBM1s-dUCCSD$T_S$ results in poor trainability\cite{truger2024warm, tilly2022variational, jattana2023improved}. However, a similar aggravation is observed for conventional dUCC ansatzes such as dUCCSD and dUCCSDT. For a more robust theoretical analysis of barren plateaus in dUCC-based ansatz, the readers are directed to the work of Mao et al. \cite{mao2023barren}

\subsection{Accuracy and Gate Utilization}
\begin{figure*}[!ht]
     \centering
 \includegraphics[width=0.9\textwidth]{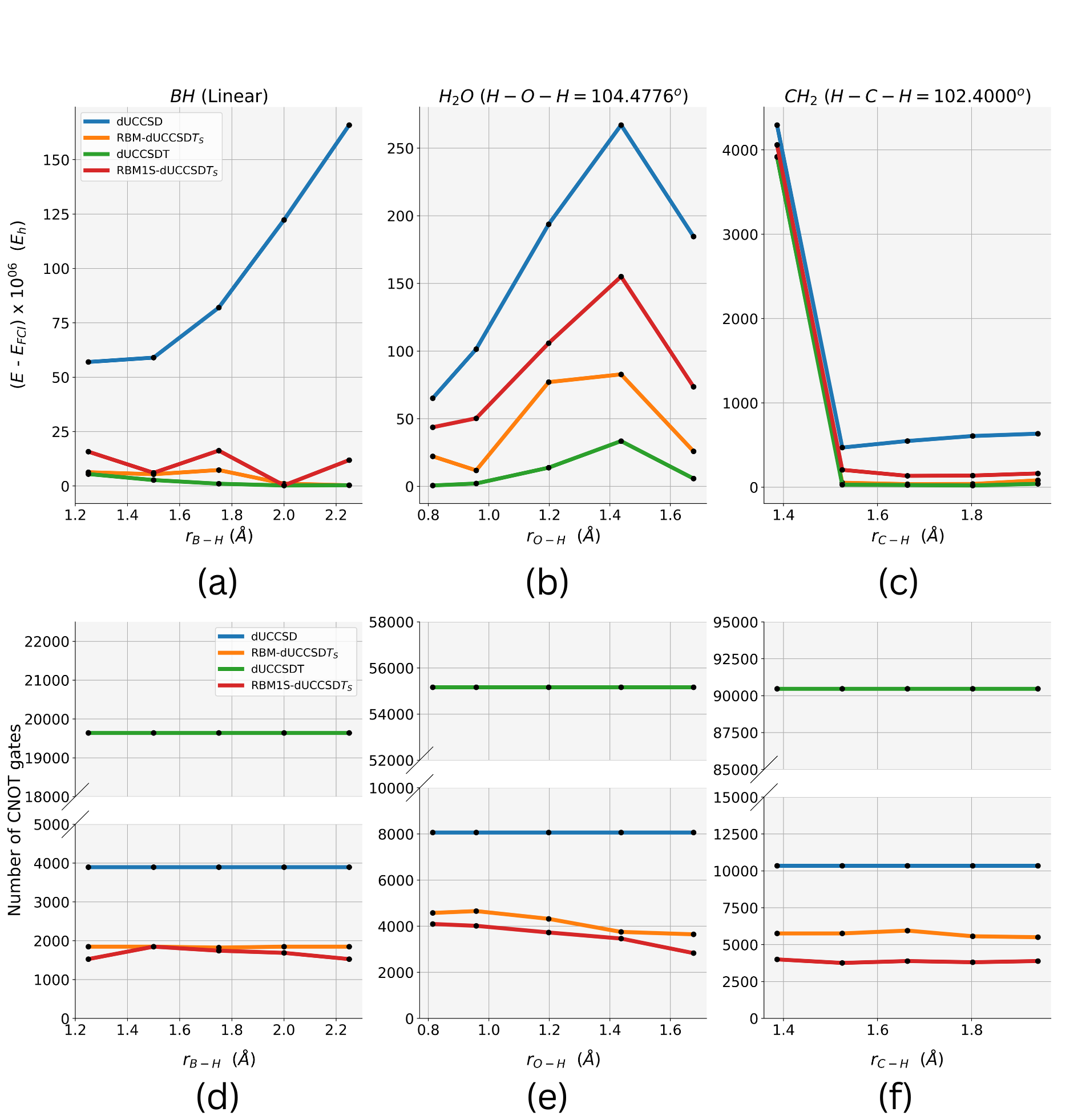}
 \caption{Energy accuracy for (a) $BH$, (b) $H_2O$ and (c) $CH_2$. Corresponding CNOT gate counts are given in (d), (e), and (f).}
     \label{fig:energy_gates}
 \end{figure*}

After conducting a rigorous analysis that establishes the efficacy of choosing zero values as initial variational parameters, we now showcase the accuracy and the number of CNOT gates associated with the RBM1s-dUCCSD$T_S$ ansatz. The results are summarized in Fig. \ref{fig:energy_gates} for three molecular systems -- (a) $BH$, (b) $H_2O$ and (c) $CH_2$. 

The variational energy minimization during Step-1 and for the final RBM1s-dUCCSD$T_S$ ansatz uses CG optimizer with maximum iterations allowed set to $10000$. The starting point for the variational optimization is set to zero for all parameters. For comparisons, Fig. \ref{fig:energy_gates} also contains results for dUCCSD, dUCCSDT, and RBM-dUCCSD$T_S$\cite{halder2024machine}. In all studied systems, the difference in energy between the RBM1s-dUCCSD$T_S$ and dUCCSDT (where all singles, doubles, and triples, which scales respectively as $\mathcal{O}(n_on_v)$, $\mathcal{O}(n_o^2n_v^2)$ and $\mathcal{O}(n_o^3n_v^3)$, $n_o$ and $n_v$ being number of occupied and unoccupied orbitals respectively, are taken) is less than $\approx2 \times 10^{-04}$. This accuracy is achieved using a substantially shallow depth ansatz (see Fig. \ref{fig:energy_gates}). The accuracy of RBM-dUCCSD$T_S$ is slightly better than RBM1s-dUCCSD$T_S$ since the latter uses RBM only once. However, the gate depth in the former is comparatively high (as much as $\approx 2 \times 10^{3}$ more in some geometries of $CH_2$). Thus, the RBM1s-dUCCSD$T_S$ strikes a good balance between the ansatz depth and accuracy with only single-step usage of RBM. Adaptive Derivative-Assembled Pseudo-Trotter ansatz Variational Quantum Eigensolver (ADAPT-VQE)\cite{grimsley2019adaptive} is a popular method that aims to construct dynamic shallow depth ansatz that can be used to calculate molecular energetics. A quantitative comparison of RBM1s-dUCCSD$T_S$ with ADAPT-VQE is provided in S5 of Supporting Information. The pool chosen for the ADAPT-VQE consists of generalized singles and doubles excitation operators. Although the former requires a slightly higher number of CNOT gates, it offers improved accuracy. Under noiseless conditions, ADAPT-VQE is among the most robust methods for constructing dynamic ansatzes. However, noise in quantum hardware can significantly impact the construction process. ADAPT-VQE relies on gradients computed on the quantum hardware to identify the most important operators from a predefined pool. Noise can distort these gradient values, potentially leading to an ansatz that fails to accurately capture correlations \cite{halder2024machine}. To address this, multiple layers of error mitigation may be required to ensure the correct operator sequence is selected. In contrast, the construction of the RBM1s-dUCC ansatz leverages generative machine learning to identify dominant operators. As demonstrated in reference \cite{halder2024machine}, such approaches are inherently more robust to noise. The shallow circuit depth of the RBM1s-dUCC ansatz, combined with its relatively noise-resilient construction process, makes it highly suitable for implementation on NISQ platforms.

\subsection{Comparative Analysis of Convergence Trajectory}

\begin{figure*}[!ht]
     \centering
 \includegraphics[width=0.9\textwidth]{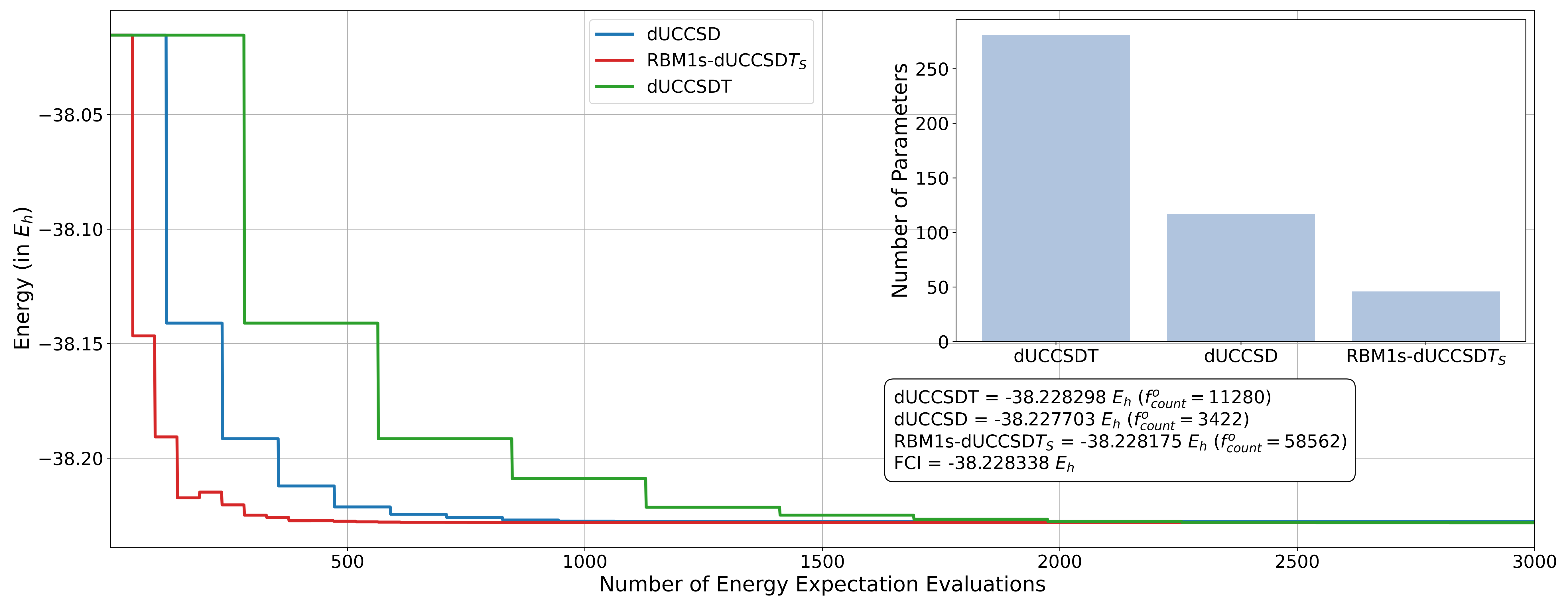}
 \caption{Convergence plot showing energy variation during optimization for $CH_2$ ($r=1.75 \times r_{eq}; \quad r_{eq}=$1.109\AA, $\angle C-H-C=102.400^o$). The X-axis denotes the optimization progress by the number of energy expectation values evaluated. The inset displays the number of optimizable parameters for the dUCCSDT, dUCCSD, and RBM1s-dUCCSD$T_S$ ansatz. The plot also indicates the final converged energies and the total number of energy expectation evaluations performed ($f^o_{count}$).}
     \label{fig:convergence}
 \end{figure*}

Current quantum hardware has limited coherence times and is plagued by other noise sources. RBM1s-dUCC ansatz overcomes both these pitfalls by being shallow in depth. Another advantage is the reduction in the number of variational parameters (as it contains fewer excitation operators). This has implications for the optimization trajectory observed during the classical energy minimization. In Fig. \ref{fig:convergence}, we provide the energy trajectory obtained using CG optimizer with maximum iterations set to $10000$ and initializing parameters with zero. In the case of RBM1s-dUCCSD$T_S$, the trajectory shows a comparatively steeper decline. This could be attributed to the smaller number of variational parameters, leading to fewer gradient evaluations. However, the long convergence \enquote{tail} associated with RBM1s-dUCCSD$T_S$ indicates the vanishing nature of gradients as the parameters reach their optimum values. As a result, the optimization proceeds slowly. Nonetheless, the accuracy remains within $\mathcal{O}(10^{-4})$ with respect to dUCCSDT. If one is restricted by the number of measurements allowed on quantum hardware, the optimization process can be stopped early. Due to the steeper decay, RBM1s-dUCCSD$T_S$ will still be able to provide accurate energy.

\section{Conclusions and Future Outlook}
In conclusion, this manuscript presents an efficient strategy that combines the Restricted Boltzmann Machine (RBM) with many-body perturbative methods to construct a compact ansatz that effectively captures a significant portion of electronic correlation. The approach starts with an easily prepared approximate wavefunction on quantum hardware. Subsequent projective measurements generate a dataset of determinants with the corresponding frequencies dictated by their coefficients in the wavefunction. Trained on this data, RBM generates new high-rank excited determinants, which are used to construct a shallow depth ansatz after being further decomposed into low-rank operators and filtered using MP2 perturbative measures. The Bayesian approach is utilized for RBM hyperparameter optimization. The shallow depth of the RBM1s-dUCC ansatz makes it highly suitable for NISQ devices. The approach depends on measurements only during the construction of the dataset. This reduces the measurement costs and makes the entire ansatz construction pipeline less prone to the detrimental effects of noise. The RBM is utilized only once and showcases an efficient use of generative machine learning for ansatz construction, particularly for cases where the machine learning model can capture the underlying correlations from an initial approximate wavefunction. The accuracy of the developed ansatz is highly dependent on the starting parameters of the variational optimization. The extensive study presented in this work provides potential initial points that lead to desired accuracy. 

A rigorous analytical study involving expressibility and trainability for the RBM1s-dUCCSD$T_S$ provides an interesting direction for future explorations. Although RBM has proved to be an effective generative model for ansatz construction, other models can be explored for better performance. Moreover, the use of generalized excitation operators (other than \textit{scatterers}) in the ansatz may be explored. The use of RBM to construct an efficient pool for other adaptive ansatz construction methods, such as ADAPT-VQE, also provides a lucrative avenue. While this work showcases the use of generative models for constructing ansatzes for fermionic systems, another interesting direction is to tackle mixed fermion-boson systems using quantum hardware based on extensions of the contracted Schr\"{o}dinger equation\cite{warren2024exact}.

\section{Supporting Information}
See the Supporting Information for - 1) Circuit implementation for dUCC based ansatz, 2) A detailed description of \textit{Scatterers}, 3) Theory and data for Bayesian hyperparameter optimization, 4) A comparison of accuracies obtained using RBM1s-dUCCSD$T_S$ ansatz when initialized from MP2 and zero values, and 5) A comparison of accuracy and CNOT gate counts between RBM1s-dUCCSD$T_S$ with ADAPT-VQE.

\section{Acknowledgment}
SH acknowledges the Council of Scientific \& Industrial Research (CSIR) for their fellowship. RM acknowledges the financial support from Industrial Research and
Consultancy Centre, IIT Bombay, and
Science and Engineering Research Board, Government
of India.

\section*{AUTHOR DECLARATIONS}
\subsection*{Conflict of Interest:}
The authors have no conflict of interest to disclose.

\section*{Data Availability}
The data is available upon reasonable request to the corresponding author.

\section*{References:}

%

\end{document}